\documentclass{article}
\usepackage{spconf,amsmath,graphicx,hyperref}
\usepackage{booktabs}
\usepackage{enumitem}
\usepackage{cite}
\usepackage{multirow}
\usepackage{arydshln}
\usepackage{tikz}
\usetikzlibrary{arrows.meta,positioning}

\title{The ISCSLP 2026 Real-World Audio-Visual Speech Enhancement Challenge: A Benchmark with Natural Mixtures and Degraded Video}
\name{\begin{tabular}{c}
Kai Li$^{1,*}$, Wenze Ren$^{2,*}$, Junjie Li$^{3,*}$, Chien-yu Huang$^{4}$, Cheng Yu$^{5}$, Peijun Yang$^{6}$, Haibin Wu$^{2}$\\
Szu-Wei Fu$^{7}$, Wen-Chin Huang$^{8}$, Hsin-Min Wang$^{9}$,
Xiaolin Hu$^{1}$, Ming Li$^{10}$, Yu Tsao$^{9,\dagger}$, DeLiang Wang$^{10}$%
\thanks{$^{*}$Equal contribution. $^{\dagger}$Corresponding author}
\end{tabular}}

\address{\begin{tabular}{c}
\textit{$^{1}$Tsinghua University $^{2}$National Taiwan University $^{3}$The Hong Kong Polytechnic University}\\
\textit{$^{4}$Carnegie Mellon University $^{5}$Ohio State University $^{6}$Wuhan University $^{7}$NVIDIA}\\
\textit{$^{8}$Nagoya University $^{9}$Academia Sinica $^{10}$The Chinese University of Hong Kong, Shenzhen}
\end{tabular}}

\begin{document}
\ninept
\maketitle

\begin{abstract}
Audio-visual speech enhancement (AVSE) uses a target speaker's visible articulation to recover that speaker's voice from overlapped speech. Yet most evaluation protocols rely on synthetic mixtures and reliable video. The ISCSLP 2026 Real-World AVSE Challenge addresses both gaps. Track~1 combines naturally recorded two-talker mixtures, which lack a clean reference, with reference-available remixes of the same speakers; Track~2 additionally degrades the target video in five ways and adds 3-m far-field recordings. Sixteen and twelve teams were ranked on speaker-disjoint test data by rank averaging over waveform fidelity, predicted quality, transcription accuracy, and speaker similarity. On Track~1 remixes, the best system reaches 12.7~dB SI-SDR and 0.85 STOI, but natural recordings remain harder: even the lowest CER rises from 9.0\% to 14.7\%. The leading systems use video mainly for speaker attribution rather than signal reconstruction, and the top two lose under 0.5~dB SI-SDR on Track~2. UTMOS and DNSMOS rank systems differently from the other metrics, so no single metric captures target-speech recovery. We release the baselines, evaluator, and official results.
\end{abstract}
\begin{keywords}
audio-visual speech enhancement, speech separation, real-world robustness, visual degradation, challenge
\end{keywords}

\vspace{-10pt}
\section{Introduction}
\label{sec:intro}

Audio-visual speech enhancement (AVSE) recovers a target speaker's voice from a mixture using the speaker's face or lip video in addition to the audio \cite{hou2018audio, sadeghi2020audio, chuang2022improved, kalkhorani2025av}. Because visible articulation is immune to acoustic interference, it provides a cue that remains informative under speaker overlap and strong noise. This principle underlies spectrogram masking, time-domain separation, and cross-modal consistency methods \cite{afouras2018conversation,wu2019time,gao2021visualvoice,ren25_avsec}. Two-stage pipelines first separate the audio and then identify the target stream by audio-visual synchronization or face-voice similarity \cite{10800033}. More recent developments include brain-inspired cross-modal architectures, intra- and inter-modality attention, and efficient time-frequency modeling \cite{li2024ctcnet,li2024iianet,pegg2024rtfsnet}. See \cite{li2025advances} for a survey.

Evaluation, however, has not kept pace with modeling. Most protocols additively mix separately recorded sources and pair the mixture with clean visual tracks, yielding a known target waveform but omitting speaker interaction, room response, ambient noise, and capture artifacts. On the acoustic side, the AVSE Challenge moved to real-world videos with added speech or noise \cite{blanco2023avsechallenge}, and the MISP 2023 Challenge turned to spontaneous far-field conversations \cite{wu2024misp}. On the visual side, the target face may in practice be occluded, blurred, frozen, misaligned, distant, missing, or of low resolution; models robust to audio-visual misalignment \cite{lee2021crossmodal} and recent systems for missing or low-quality video \cite{pan2024ravss,chen2025hpcnet,11561071,11209435} address such conditions, yet no benchmark has treated visual corruption as a controlled variable. No reproducible protocol has combined the two, evaluating naturally mixed audio with unreliable visual evidence; the REAL-TSE Challenge \cite{realtse_challenge} is closest in spirit but conditions on an enrollment utterance rather than video.


The ISCSLP 2026 Real-World AVSE Challenge addresses this gap. Given a monaural mixture and the target speaker's video, systems are evaluated on naturally recorded two-talker mixtures alongside reference-available remixes of the same speakers (Track 1) and under controlled degradation of the target video (Track 2). The official results yield three findings: (i) natural recordings remain harder than remixes of the same speakers, even for the best systems; (ii) the leading systems use video primarily for speaker attribution rather than signal reconstruction; and (iii) non-intrusive quality predictors rank systems differently from fidelity, transcription, and speaker-similarity metrics. The protocol, evaluator, and results are available on the challenge website\footnote{\url{https://real-world-avse.github.io/}} and in the baseline repository\footnote{\url{https://github.com/Real-World-AVSE/Baseline}}.

\vspace{-8pt}
\section{Challenge Description}
\subsection{Task Definition and Tracks}
Let $x(t)$ denote the observed monaural mixture and $v_k(\tau)$ the visual stream of the target speaker $k$, where $t$ and $\tau$ index audio and video time, respectively. The system estimates the target waveform as
\begin{equation}
    \hat{s}_k(t) = f_\theta\bigl(x(t), v_k(\tau)\bigr),
    \label{eq:task}
\end{equation}
where $f_\theta$ is the AVSE function with trainable parameters $\theta$.

The benchmark comprises two tracks:
\begin{itemize}
  \item \textbf{Track 1: Real-World Mixed Scenarios.} The principal condition is naturally recorded two-talker speech, which preserves the joint effects of overlap, room acoustics, ambient sound, and recording chain. Reference-available remixes of the same speakers are included as a controlled anchor.
  \item \textbf{Track 2: Visual Degradation.} The target video of the Track~1 clips is degraded in five ways while the audio is left unchanged, and additional 3-m far-field recordings from the same sessions are added (Sec.~\ref{sec:data}).
\end{itemize}

\vspace{-4pt}
\subsection{Dataset Description}
\label{sec:data}


\begin{table}[t]
\caption{Released challenge data. Mix and Remix are clip counts per scene, and each clip yields two target items, one per speaker. Track~2 also includes the far-field recordings.}
\label{tab:data-size}
\centering
\scriptsize
\begin{tabular}{l|rrr|rrr}
\toprule
& \multicolumn{3}{c|}{\textbf{Development}} & \multicolumn{3}{c}{\textbf{Test}}\\
\cmidrule(lr){2-4} \cmidrule(lr){5-7}
\textbf{Track} & \textbf{Mix} & \textbf{Remix} & \textbf{Items} & \textbf{Mix} & \textbf{Remix} & \textbf{Items}\\
\midrule
1 & 1,242 & 900   & 4,284 & 2,472 & 1,785 & 8,514\\
2 & 1,527 & 1,098 & 5,250 & 2,820 & 2,121 & 9,882\\
\bottomrule
\end{tabular}
\vspace{-6mm}
\end{table}
\begin{figure}[t]
\centering
\includegraphics[width=1.05\columnwidth]{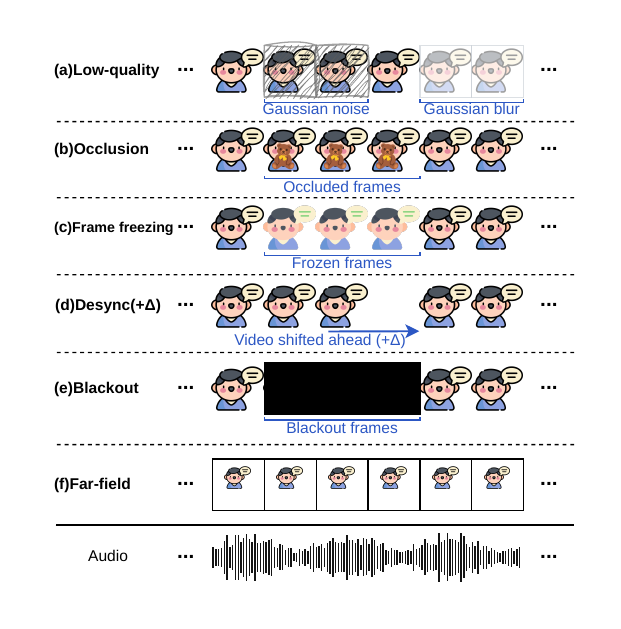}
\vspace{-6mm}
\caption{Track~2 conditions. In (a)--(e), only the target-speaker video $v_k(\tau)$ in Eq.~(\ref{eq:task}) is perturbed, offline, while the audio $x(t)$ (bottom row) stays unchanged; in (d), the video may lead ($+\Delta$, shown) or lag ($-\Delta$) the audio. Condition (f) is a real far-field recording at about 3~m, which changes both $x(t)$ and $v_k(\tau)$.}
\label{fig:visual_perturbation}
\end{figure}

\vspace{-4pt}


Table~\ref{tab:data-size} lists the released clips and target items \cite{yang2026identity}. The audio is 16-kHz mono; face videos are 256$\times$256 pixels at 25~fps with optional facial-landmark files. The corpus contains seven groups of two speakers, three for development and four for test, so the splits are speaker-disjoint. Each split has two scenes. In the \emph{mix} scene, both speakers are recorded together and no isolated target exists. In the \emph{remix} scene, two separately recorded single-speaker segments are added at an SNR sampled from $-5$ to $5$~dB, and the target source is kept as a reference. Remix thus supports clean-waveform metrics, whereas mix retains the natural recording condition and is scored with reference-free metrics only (Sec.~\ref{sec:evaluation}); gains on remix should not come at the expense of mix.

Fig.~\ref{fig:visual_perturbation} illustrates the six Track~2 conditions: five perturbations\footnote{\url{https://github.com/Real-World-AVSE/Baseline/tree/main/look2hear/datas/visual_perturb}} of the target-speaker video that leave the audio unchanged, (a)--(e), and one additional recording condition, (f):
\begin{itemize}[noitemsep]
  \item \emph{Low-quality video}: Gaussian noise or blur applied to a contiguous segment.
  \item \emph{Occlusion}: a random object image overlaid on a facial landmark region, typically the mouth.
  \item \emph{Frame freezing}: one frame repeated over an interval while the audio continues.
  \item \emph{Audio-video desynchronization}: the video shifted ahead of or behind the audio, with the frame count preserved.
  \item \emph{Blackout}: a contiguous block of frames replaced with all-black frames.
  \item \emph{Far-field recording}: additional real recordings at approximately 3~m; unlike the above, this is not an offline video modification and alters both modalities.
\end{itemize}

\vspace{-9pt}

\subsection{Evaluation Protocol}
\label{sec:evaluation}

\begin{table}[t]
\caption{Evaluation metrics: required reference, scenes scored, and backend.}
\label{tab:metrics}
\centering
\scriptsize
\setlength{\tabcolsep}{3.0pt}
\begin{tabular}{p{0.14\linewidth}p{0.30\linewidth}p{0.20\linewidth}p{0.18\linewidth}}
\toprule
\textbf{Metric} & \textbf{Reference} & \textbf{Scene} & \textbf{Backend}\\
\midrule
SI-SDR  & Clean target waveform & Remix      & torchmetrics\\
PESQ    & Clean target waveform & Remix      & torchmetrics\\
STOI    & Clean target waveform & Remix      & torchmetrics\\
UTMOS   & No                    & Remix\&Mix & UTMOSv2\\
DNSMOS  & No                    & Remix\&Mix & ONNX\\
CER     & Transcript            & Remix\&Mix & Fun-ASR\\
SPK-SIM & Enrollment voiceprint & Remix\&Mix & WeSpeaker\\
\bottomrule
\end{tabular}
\end{table}
Table~\ref{tab:metrics} lists the metrics and the reference each requires. SI-SDR~\cite{leroux2019sdr}, PESQ~\cite{pesq}, and STOI~\cite{stoi} need the clean target and are therefore computed on the \emph{remix} scene only. UTMOSv2~\cite{utmosv2} and DNSMOS~\cite{dnsmos} predict quality without a reference, Fun-ASR \cite{funasr} provides character error rate (CER) against organizer-provided transcripts, and WeSpeaker embeddings provide enrollment-based speaker similarity \cite{wespeaker}; these four are computed on both scenes. All summary metrics are arithmetic means over target items; speaker similarity is the cosine similarity between the enhanced-speech embedding and a per-speaker centroid built from clean remix sources. The official leaderboard score is
\begin{equation}
  \mathrm{OVRL}=\frac{1}{|\mathcal{M}|}\sum_{m\in\mathcal{M}} r_m,
  \label{eq:ovrl}
\end{equation}
where $r_m$ is the rank of a team on metric $m$ within the selected scope, with CER ranked in ascending order and all other metrics in descending order. DNS-P808, DNS-SIG, and DNS-BAK are displayed for reference\footnote{\url{https://avse.renwenze.com/}} but excluded from the average; only DNS-OVRL enters it.

\vspace{-4pt}
\section{Baseline system and Participants}

\subsection{Baseline system}
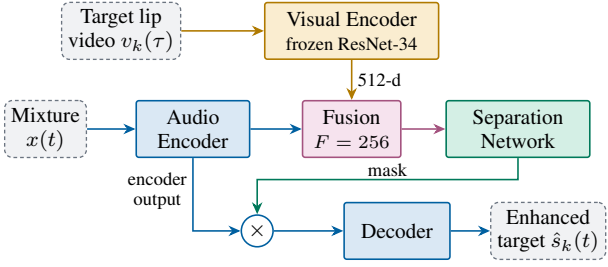
\begin{figure}[t]
\centering
\definecolor{avBlue}{HTML}{0060A0}    \definecolor{avBlueBg}{HTML}{DAE8F4}
\definecolor{avOrange}{HTML}{B07A00}  \definecolor{avOrangeBg}{HTML}{FAEDD0}
\definecolor{avPurple}{HTML}{A4537F}  \definecolor{avPurpleBg}{HTML}{F6E3EE}
\definecolor{avGreen}{HTML}{00795A}   \definecolor{avGreenBg}{HTML}{D4EFE5}
\definecolor{avGrey}{HTML}{6B7480}    \definecolor{avGreyBg}{HTML}{F1F2F4}
\begin{tikzpicture}[
  font=\footnotesize,
  >={Stealth[length=1.6mm,width=1.4mm]},
  block/.style   = {draw, line width=.6pt, rounded corners=1pt, align=center,
                    inner sep=2pt, minimum height=7.5mm},
  audio/.style   = {block, draw=avBlue,   fill=avBlueBg},
  visual/.style  = {block, draw=avOrange, fill=avOrangeBg},
  fusion/.style  = {block, draw=avPurple, fill=avPurpleBg},
  sepnet/.style  = {block, draw=avGreen,  fill=avGreenBg},
  io/.style      = {draw=avGrey, line width=.5pt,
                    dash pattern=on 1.4pt off 1.1pt, rounded corners=3pt,
                    fill=avGreyBg, align=center, inner sep=2.5pt,
                    minimum height=7.5mm},
  op/.style      = {draw=avBlue, line width=.6pt, fill=white, circle,
                    inner sep=0pt, minimum size=4.2mm},
  lbl/.style     = {font=\scriptsize, align=center, inner sep=1pt},
  aflow/.style   = {->, line width=.6pt, draw=avBlue},
  vflow/.style   = {->, line width=.6pt, draw=avOrange},
  pflow/.style   = {->, line width=.6pt, draw=avPurple},
  mflow/.style   = {->, line width=.6pt, draw=avGreen},
]

\node[io]                          (inv)  at (1.70,3.35) {Target lip\\video $v_k(\tau)$};
\node[visual, minimum width=23mm]  (encv) at (4.75,3.35) {Visual Encoder\\\scriptsize frozen ResNet-34};

\node[io]                          (inx)  at (0.70,2.05) {Mixture\\$x(t)$};
\node[audio,  minimum width=15mm]  (enca) at (2.65,2.05) {Audio\\Encoder};
\node[fusion, minimum width=13mm]  (fus)  at (4.75,2.05) {Fusion\\\scriptsize $F=256$};
\node[sepnet, minimum width=19mm]  (sep)  at (6.95,2.05) {Separation\\Network};

\node[op]                          (mul)  at (3.50,0.70) {$\times$};
\node[audio, minimum width=14mm]   (dec)  at (5.35,0.70) {Decoder};
\node[io]                          (out)  at (7.35,0.70) {Enhanced\\target $\hat{s}_k(t)$};

\draw[vflow] (inv)  -- (encv);
\draw[vflow] (encv) -- node[lbl, right=1pt] {512-d} (fus);
\draw[aflow] (inx)  -- (enca);
\draw[aflow] (enca) -- (fus);
\draw[pflow] (fus)  -- (sep);

\draw[mflow] (sep.south) -- (6.95,1.38)
             -- node[lbl, above] {mask} (3.50,1.38) -- (mul.north);

\draw[aflow] (enca.south)
             -- node[lbl, left=1pt, pos=0.45] {encoder\\output} (2.65,0.70)
             -- (mul.west);

\draw[aflow] (mul) -- (dec);
\draw[aflow] (dec) -- (out);

\end{tikzpicture}
\caption{Official AV-ConvTasNet baseline architecture.}
\vspace{-3mm}
\label{fig:official-backbone}
\end{figure}


The official baseline is AV-ConvTasNet \cite{wu2019time}, a two-stream target-speaker separator that extends Conv-TasNet \cite{convtasnet} with a visual conditioning branch (Fig.~\ref{fig:official-backbone}). The audio branch retains the Conv-TasNet design: a learned filterbank encodes the mixture waveform into a latent representation, a non-causal temporal convolutional separator estimates a mask over this representation, and a learned decoder reconstructs the target waveform from the masked representation. The visual branch is a frozen ResNet-34 lip-reading encoder with a 3D convolutional stem that maps normalized $88\times88$ lip crops of the target speaker to 512-dimensional features. We concatenate these features with the audio-branch bottleneck features and project them to a fusion dimension of $F=256$, so the mask is estimated from both modalities. All other hyperparameters follow the released repository.

We generate training mixtures on the fly. For Track~1, two single-speaker VoxCeleb2 utterances \cite{voxceleb2} are combined into a two-speaker remix at each iteration. For Track~2, we retain the same audio pipeline and additionally subject the target video to online visual perturbations, namely occlusion, low resolution, frame freezing, dropped frames, and audio-visual desynchronization. In both cases, the model is trained with a negative SI-SDR objective on the target speaker \cite{leroux2019sdr} using Adam at an initial learning rate of $10^{-3}$ with a ReduceLROnPlateau scheduler, and a separate checkpoint is released for each track. Because the training mixtures are simulated from VoxCeleb2 rather than drawn from the challenge recordings, this recipe should be treated as a reproducible starting point rather than a simulation of the official evaluation distribution.

\vspace{-6pt}
\subsection{Participants}


The challenge attracted registrations from 51 teams. Of these, 16 teams from both academia and industry submitted enhanced audio to the Track~1 leaderboard, and 12 of them also submitted to the Track~2 leaderboard.
Each team is represented by its best entry, i.e., the submission with the lowest OVRL, and the complete leaderboard is available on the challenge website. Below, we describe the three highest-ranked systems that submitted system descriptions, namely audioman, AITD, and Yijiahe Robot; twilight, which ranks third on Track~1 and fourth on Track~2, did not submit a description and is therefore not covered. Sec.~\ref{sec:results} compares these systems with the remaining entries and the official baseline.

\vspace{-6pt}
\subsection{Top-Ranked Systems}
\label{sec:top_systems}
All three systems replace the time-domain baseline with time-frequency backbones, train on simulated reverberant and noisy mixtures, and optimize metric-oriented objectives beyond SI-SDR; they differ mainly in how they use video.

\noindent\textbf{AITD.}
Video never enters signal reconstruction. A blind STFT-domain dual-path Transformer, trained on audio-only corpora, first with PIT SI-SNR and then with perceptual (UTMOS, PESQ, WavLM \cite{Chen_2022}) losses, and finally fine-tuned on the Track~1 development set, produces two anonymous streams. An AVSelector built on partially fine-tuned auto-AVSR \cite{Ma_2023} encoders scores each stream against the lip video and assigns identities by bipartite matching; visual augmentation during its training makes the assignment robust to degraded video.

\noindent\textbf{audioman.}
The only fully audio-visual separator among the three described systems fuses AV-HuBERT \cite{shi2022learning} features of both speakers into a SIR-progressive TF-GridNet that emits two coarse estimates, one per input video, with only competing speech removed. A second TF-GridNet \cite{10214650} then refines both from the mixture under reconstruction, over-suppression, differentiable PESQ, and adversarial losses. Because refinement trades waveform fidelity for predicted quality, the coarse and refined estimates are linearly blended with a development-selected weight. Training uses VoxCeleb2 and MISP~2023 with RIRs and noise, plus visual degradation and modality dropout for Track~2.

\noindent\textbf{Yijiahe Robot.}
DAVE separates with an audio-only TIGER \cite{xu2025tiger} backbone and, like AITD, uses video only for attribution, through a weighted vote over voiceprint similarity, lip-sync detectors, and lip-keypoint motion. It is trained on a large simulated corpus derived from meeting recordings, official-protocol remixes of the development set, and EchoSet \cite{xu2025tiger}, with PIT SI-SDR plus differentiable surrogates for CER, speaker similarity, STOI, PESQ, and UTMOSv2. A scene classifier applies GAN denoising and loudness normalization only to clips detected as natural mixtures, leaving the remix scene untouched.

\vspace{-6pt}
\section{Results and Analysis}
\label{sec:results}

\begin{figure}[t]
\centering
\includegraphics[width=1.05\columnwidth]{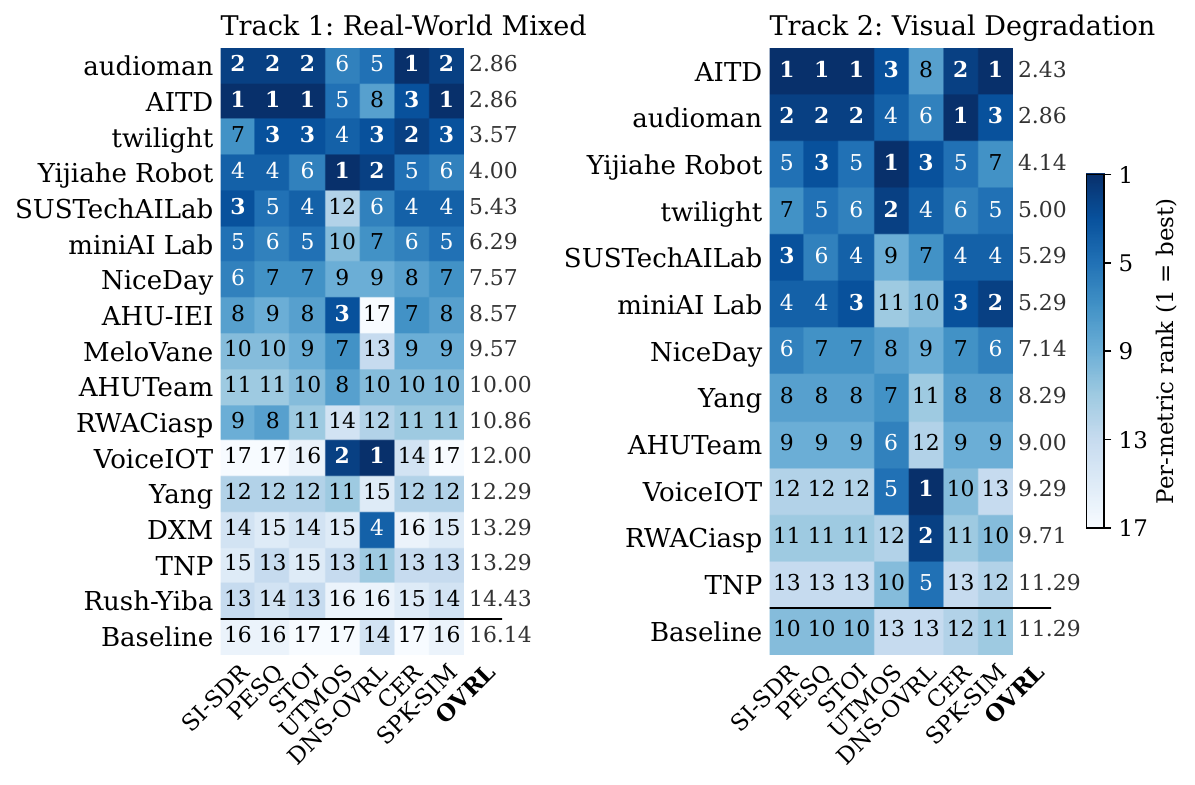}
\caption{Official test-set leaderboard. Each cell shows an entry's rank on one metric, and OVRL averages the seven ranks as in Eq.~\eqref{eq:ovrl}, so lower is better. The baseline in the bottom row is ranked together with the teams.}
\vspace{-3mm}
\label{fig:rank heatmap}
\end{figure}


\begin{table*}[t]
\centering
\caption{Test-set results of the three described systems and the official baseline on Track~1 (Real-World Mixed) and Track~2 (Visual Degradation). The mix scene has no clean reference, so SI-SDR (dB), PESQ, and STOI are available on the remix scene only. Best value per column within each track in bold.}
\label{tab:results}
\footnotesize
\setlength{\tabcolsep}{3.5pt}
\begin{tabular}{c l cccc@{\hspace{1.5em}}ccccccc}
\toprule
& & \multicolumn{4}{c}{Mix scene (no clean reference)} & \multicolumn{7}{c}{Remix scene (clean reference available)} \\
\cmidrule(lr){3-6} \cmidrule(lr){7-13}
& System & UTMOS$\uparrow$ & DNS-OVRL$\uparrow$ & CER$\downarrow$ & SPK-SIM$\uparrow$ & SI-SDR$\uparrow$ & PESQ$\uparrow$ & STOI$\uparrow$ & UTMOS$\uparrow$ & DNS-OVRL$\uparrow$ & CER$\downarrow$ & SPK-SIM$\uparrow$ \\
\midrule
\multirow{4}{*}{\rotatebox[origin=c]{90}{\textbf{Track 1}}}
& audioman      & 2.024          & 1.802          & \textbf{0.147} & 0.734          & 10.699          & 2.933          & 0.841          & 2.143          & \textbf{1.874} & \textbf{0.090} & 0.771 \\
& AITD          & 2.072          & 1.664          & 0.183          & \textbf{0.751} & \textbf{12.716} & \textbf{3.000} & \textbf{0.852} & 2.139          & 1.742          & 0.092          & \textbf{0.794} \\
& Yijiahe Robot & \textbf{2.652} & \textbf{2.225} & 0.189          & 0.696          & 10.232          & 2.717          & 0.817          & \textbf{2.921} & 1.742          & 0.146          & 0.768 \\
\cmidrule(lr){2-13}
& Baseline      & 0.730          & 1.467          & 1.059          & 0.334          & $-5.925$        & 1.137          & 0.304          & 0.926          & 1.420          & 0.979          & 0.320 \\
\midrule
\multirow{4}{*}{\rotatebox[origin=c]{90}{\textbf{Track 2}}}
& audioman      & 2.045          & 1.731          & \textbf{0.175} & 0.728          & 10.223          & 2.888          & 0.830          & 2.159          & \textbf{1.798} & \textbf{0.112} & 0.766 \\
& AITD          & 2.092          & 1.638          & 0.183          & \textbf{0.748} & \textbf{12.255} & \textbf{2.939} & \textbf{0.844} & 2.178          & 1.713          & 0.114          & \textbf{0.785} \\
& Yijiahe Robot & \textbf{2.648} & \textbf{2.241} & 0.234          & 0.678          & 8.933           & 2.615          & 0.789          & \textbf{2.922} & 1.708          & 0.201          & 0.745 \\
\cmidrule(lr){2-13}
& Baseline      & 1.097          & 1.166          & 1.094          & 0.401          & $-1.689$        & 1.304          & 0.502          & 1.104          & 1.262          & 0.998          & 0.389 \\
\bottomrule
\end{tabular}
\end{table*}

\subsection{Overall Results}
\label{ssec:overall}
Fig.~\ref{fig:rank heatmap} shows the per-metric ranks and the resulting OVRL score of every entry, and Table~\ref{tab:results} shows the absolute scores of the three systems described in Sec.~\ref{sec:top_systems} and the official baseline. The same four teams occupy the first four positions on both tracks: audioman and AITD tie on Track~1 (OVRL 2.86), and AITD leads Track~2 (2.43) ahead of audioman (2.86). The baseline, trained only on simulated VoxCeleb2 mixtures, does not transfer to the challenge recordings: it ranks last on Track~1 and ties for last on Track~2, with a remix SI-SDR of $-5.925$~dB and a CER above 1.0, indicating that the total number of ASR errors exceeds the number of reference characters.

By contrast, AITD reaches 12.72~dB SI-SDR and 0.85 STOI on the Track~1 remix scene, and audioman achieves the lowest CER on both scenes of both tracks: 9.0\% on remix and 14.7\% on mix for Track~1, and 11.2\% and 17.5\% for Track~2.

Natural recordings nevertheless remain harder than remixes of the same speakers. Across all three systems and both tracks, CER is higher, and SPK-SIM is lower on the mix scene than on the remix scene; on Track~1, AITD's CER doubles from 9.2\% to 18.3\%. The mix and remix clips are different recordings, so this comparison is not exact, but the direction is the same for every system on both tracks. Even the best system still makes about one error in every seven characters on naturally recorded overlap (14.7\% CER on Track~1), so the real-world condition is far from solved.
\vspace{-6pt}
\subsection{Role of Video under Visual Degradation}
\label{ssec:video}
The leading systems use video mainly to decide which output belongs to the target, not to reconstruct the signal. AITD and Yijiahe Robot separate the two voices with an audio-only model and use the video only afterwards, to pick the target speaker's output. AITD, in which video never enters reconstruction, ranks first among all teams on SI-SDR, PESQ, STOI, and SPK-SIM on both tracks. Audioman, the only one of the three that fuses visual features into the separator, leads on CER. Thus, with two talkers, a strong audio-only separator and a reliable visual selector are enough to achieve the best fidelity scores. This does not show that audio-visual fusion is worse, since the systems also differ in training data and loss functions.

Visual degradation therefore costs little. From Track~1 to Track~2, SI-SDR decreases by 0.46~dB for AITD and 0.48~dB for audioman, STOI decreases by less than 0.012, and CER increases by 0.8 and 2.5 points; Yijiahe Robot loses more ($-1.30$~dB SI-SDR, $+4.9$ CER points). When the video fails, the main risk is a wrong speaker assignment, not worse separation, and both systems prepare for it in training: AITD augments the video seen by
its selector, and audioman applies modality dropout and visual corruption.

This comparison is not a paired ablation: the Track~2 test set is larger (Table~\ref{tab:data-size}), its far-field items change both the audio and the video, and teams could submit track-specific systems. For example, the baseline scores 4.2~dB higher in SI-SDR on Track~2 than on Track~1, which reflects the different checkpoints and test items, not a benefit of degraded video. With this caveat, Track~2 mainly tests whether a system still picks the right speaker when the video is degraded, not whether video still helps separation. Whether video helps separation under more challenging conditions, such as those involving more talkers or similar voices, remains an open question.

\vspace{-4pt}
\subsection{Disagreement among Metrics}
\label{ssec:metrics}
Fig.~\ref{fig:rank heatmap} shows that SI-SDR, PESQ, STOI, CER, and SPK-SIM rank the teams in nearly the same order: AITD and audioman, for example, are in the top three on all five metrics on both tracks. UTMOS and DNS-OVRL rank the teams differently. On Track~1, VoiceIOT ranks first on DNS-OVRL and second on UTMOS but last, below the baseline, on SI-SDR, PESQ, and SPK-SIM; on Track~2, VoiceIOT and RWACiasp hold the top two DNS-OVRL positions while ranking eleventh or twelfth on SI-SDR, PESQ, and STOI. The two predictors also disagree: AHU-IEI ranks third on UTMOS and last on DNS-OVRL on Track~1.

Non-intrusive predictors reward outputs that sound clean whether or not the target's content and identity are preserved, so aggressive suppression or resynthesis can raise them. For example, on the Track~1 mix scene, the baseline's DNS-OVRL is within 0.2 of AITD's (1.47 vs.\ 1.66), although its output is untranscribable (CER 106\% vs.\ 18\%). The described systems show the same effect: Yijiahe Robot applies GAN denoising only to clips detected as natural mixtures, and only on the mix scene is its DNS-OVRL higher than those of audioman and AITD (2.23 vs.\ 1.80 and 1.66 on Track~1), even though its CER is the highest and its SPK-SIM the lowest of the three.

This matters most in the mix scene, which has no clean reference: there, UTMOS and DNS-OVRL should be interpreted only alongside CER and SPK-SIM, which check that the target's content and identity are preserved. The rank average limits the influence of any single metric: VoiceIOT finishes twelfth on Track~1 despite two top-two placements.

\vspace{-9pt}
\section{Conclusion}
We presented the ISCSLP 2026 Real-World AVSE Challenge, to our knowledge the first reproducible benchmark that evaluates AVSE on naturally recorded two-talker mixtures and under controlled degradation of the target video. It provides speaker-disjoint data, a rank-averaged score over fidelity, predicted quality, transcription, and speaker similarity, and a public baseline, evaluator, and leaderboard. Results from 16 and 12 teams yield three findings. First, natural recordings remain harder than remixes of the same speakers: even for the best system, CER rises from 9.0\% on remixes to 14.7\% on natural recordings. Second, the leading systems use video mainly for speaker attribution rather than signal reconstruction, and the two top-ranked systems lose less than 0.5~dB SI-SDR from Track~1 to Track~2. Third, UTMOS and DNSMOS rank systems differently from the other metrics, so predicted quality should be interpreted alongside CER and speaker similarity, especially when no clean reference exists.

In future editions, we plan to add more talkers and similar voices, in conditions where visual selection alone may not suffice. We also plan to pair clean and degraded versions of the same clips so each degradation can be measured directly, and to run listening tests to calibrate the learned quality predictors.

\section{Generative AI Use Disclosure}

Generative AI was used only for language polishing; the authors take full responsibility for all content. AI tools did not contribute to the substantive scientific content.

\bibliographystyle{IEEEbib}
\bibliography{refs}

@article{hou2018audio,
  title={Audio-visual speech enhancement using multimodal deep convolutional neural networks},
  author={Hou, Jen-Cheng and others},
  journal={IEEE Trans. on Emerging Topics in Computational Intelligence},
  year={2018},
  publisher={IEEE}
}

@article{sadeghi2020audio,
  title={Audio-visual speech enhancement using conditional variational auto-encoders},
  author={Sadeghi, Mostafa and others},
  journal={IEEE Trans. Audio, Speech, Language Process.},
  year={2020},
  publisher={IEEE}
}

@ARTICLE{kalkhorani2025av,
  author={Kalkhorani and others},
  journal={IEEE Journal of Selected Topics in Signal Processing}, 
  title={{AV-CrossNet}: An Audiovisual Complex Spectral Mapping Network for Speech Separation by Leveraging Narrow- and Cross-Band Modeling}, 
  year={2025}}

@article{chuang2022improved,
  title={Improved lite audio-visual speech enhancement},
  author={Chuang, Shang-Yi and others},
  journal={IEEE Trans. Audio, Speech, Language Process.},
  year={2022}
}

@inproceedings{ren25_avsec,
  title     = {{BAV-MossFormer2}: Enhanced {MossFormer}2 for Binaural Audio-Visual Speech Enhancement},
  author    = {Wenze Ren and Kai Li and others},
  year      = {2025},
  booktitle = {4th Cogmhear AVSEC}
}

@INPROCEEDINGS{10800033,
  author={Ren, Wenze and others},
  booktitle={Proc. O-COCOSDA}, 
  title={Robust Audio-Visual Speech Enhancement: Correcting Misassignments in Complex Environments With Advanced Post-Processing}, 
  year={2024},
}

@article{realtse_challenge,
  author        = {Wang, Shuai and others},
  title         = {{SLT 2026 REAL-TSE Challenge}: Real-world Target Speaker Extraction from Conversational Recordings},
  year          = {2026}
}

@article{convtasnet,
  author  = {Luo, Yi and Mesgarani, Nima},
  title   = {{Conv-TasNet}: Surpassing Ideal Time-Frequency Magnitude Masking for Speech Separation},
  journal = {IEEE Trans. Audio, Speech, Language Process.},
  year    = {2019}
}

@inproceedings{voxceleb2,
  author    = {Chung, Joon Son and others},
  title     = {{VoxCeleb2}: Deep Speaker Recognition},
  booktitle = {Proc. Interspeech},
  year      = {2018},
}

@INPROCEEDINGS{pesq,
  author={Rix, A.W. and others},
  booktitle={Proc. ICASSP}, 
  title={Perceptual evaluation of speech quality ({PESQ})-a new method for speech quality assessment of telephone networks and codecs}, 
  year={2001}
 }

@article{stoi,
  author  = {Taal, Cees H. and others},
  title   = {An Algorithm for Intelligibility Prediction of Time-Frequency Weighted Noisy Speech},
  journal = {IEEE Trans. Audio, Speech, Language Process.},
  year    = {2011}
}

@inproceedings{wespeaker,
  author    = {Wang, Hongji and others},
  title     = {{WeSpeaker}: A Research and Production Oriented Speaker Embedding Learning Toolkit},
  booktitle = {Proc. ICASSP},
  year      = {2023}
}

@inproceedings{utmosv2,
  author    = {Baba, Kaito and others},
  title     = {The {T05} System for The {VoiceMOS} Challenge 2024: Transfer Learning from Deep Image Classifier to Naturalness {MOS} Prediction of High-Quality Synthetic Speech},
  booktitle = {Proc. SLT},
  year      = {2024}
}

@INPROCEEDINGS{dnsmos,
  author={Reddy, Chandan K A and others},
  booktitle={Proc. ICASSP}, 
  title={Dnsmos P.835: A Non-Intrusive Perceptual Objective Speech Quality Metric to Evaluate Noise Suppressors}, 
  year={2022}
}

@article{funasr,
  author       = {An, Keyu and others},
  title        = {{Fun-ASR} Technical Report},
  journal      = {arXiv preprint arXiv:2509.12508},
  year         = {2025}
}

@INPROCEEDINGS{11209435,
  author={Li, Junjie and others},
  booktitle={Proc. ICME}, 
  title={{MoMuSE}: Momentum Multi-modal Target Speaker Extraction for Real-time Scenarios with Impaired Visual Cues}, 
  year={2025}}

@ARTICLE{11561071,
  author={Li, Junjie and others},
  journal={IEEE Trans. Audio, Speech, Language Process.}, 
  title={{MeMo}: Attentional Momentum for Real-Time Audio-Visual Target Speaker Extraction Under Impaired Visual Conditions}, 
  year={2026}}

@inproceedings{wu2019time,
  title={Time domain audio visual speech separation},
  author={Wu, Jian and others},
  booktitle={Proc. ASRU},
  year={2019}
}

@inproceedings{afouras2018conversation,
  author    = {Afouras, Triantafyllos and others},
  title     = {The Conversation: Deep Audio-Visual Speech Enhancement},
  booktitle = {Proc. Interspeech},
  year      = {2018}
}

@inproceedings{gao2021visualvoice,
  author    = {Gao, Ruohan and Grauman, Kristen},
  title     = {{VisualVoice}: Audio-Visual Speech Separation with Cross-Modal Consistency},
  booktitle = {Proc. CVPR},
  year      = {2021},
}

@inproceedings{li2024iianet,
  author    = {Li, Kai and others},
  title     = {{IIANet}: An Intra- and Inter-Modality Attention Network for Audio-Visual Speech Separation},
  booktitle = {Proc. ICML},
  year      = {2024}
}

@inproceedings{pegg2024rtfsnet,
  author    = {Pegg, Samuel and others},
  title     = {{RTFS-Net}: Recurrent Time-Frequency Modelling for Efficient Audio-Visual Speech Separation},
  booktitle = {Proc. ICLR},
  year      = {2024}
}

@article{li2024ctcnet,
  author  = {Li, Kai and others},
  title   = {An Audio-Visual Speech Separation Model Inspired by Cortico-Thalamo-Cortical Circuits},
  journal = {IEEE Trans. on Pattern Analysis and Machine Intelligence},
  year    = {2024}
}

@article{li2025advances,
  author  = {Li, Kai and others},
  title   = {Advances in Speech Separation: Techniques, Challenges, and Future Trends},
  journal = {arXiv preprint arXiv:2508.10830},
  year    = {2025}
}

@inproceedings{blanco2023avsechallenge,
  author    = {Blanco, Andrea Lorena Aldana and others},
  title     = {{AVSE Challenge}: Audio-Visual Speech Enhancement Challenge},
  booktitle = {Proc. SLT},
  year      = {2023}
}

@inproceedings{wu2024misp,
  author    = {Wu, Shilong and others},
  title     = {The Multimodal Information Based Speech Processing ({MISP}) 2023 Challenge: Audio-Visual Target Speaker Extraction},
  booktitle = {Proc. ICASSP},
  year      = {2024}
}

@INPROCEEDINGS{lee2021crossmodal,
  author={Lee, Jiyoung and others},
  booktitle={Proc. CVPR}, 
  title={Looking into Your Speech: Learning Cross-modal Affinity for Audio-visual Speech Separation}, 
  year={2021}}

@inproceedings{pan2024ravss,
  author    = {Pan, Tianrui and others},
  title     = {{RAVSS}: Robust Audio-Visual Speech Separation in Multi-Speaker Scenarios with Missing Visual Cues},
  booktitle = {Proc. ACM Multimedia},
  year      = {2024}
}

@article{chen2025hpcnet,
  author  = {Chen, Hang and others},
  title   = {{HPCNet}: Hybrid Pixel and Contour Network for Audio-Visual Speech Enhancement With Low-Quality Video},
  journal = {IEEE Journal of Selected Topics in Signal Processing},
  year    = {2025},
}

@inproceedings{leroux2019sdr,
  author    = {Le Roux, Jonathan and others},
  title     = {{SDR}--Half-Baked or Well Done?},
  booktitle = {Proc. ICASSP},
  year      = {2019}
}

@article{yang2026identity,
  title={Identity-Faithful Audio-Visual Target Speaker Extraction with {QIANGDA} and {VOXBLINK2-AVSE}},
  author={Yang, Peijun and others},
  year={2026}
}

@article{Chen_2022,
   title={WavLM: Large-Scale Self-Supervised Pre-Training for Full Stack Speech Processing},
   journal={IEEE Journal of Selected Topics in Signal Processing},
   author={Chen, Sanyuan and others},
   year={2022}}

@inproceedings{Ma_2023,
   title={Auto-AVSR: Audio-Visual Speech Recognition with Automatic Labels},
   booktitle={Proc. ICASSP},
   publisher={IEEE},
   author={Ma, Pingchuan and others},
   year={2023}}

@inproceedings{
shi2022learning,
title={Learning Audio-Visual Speech Representation by Masked Multimodal Cluster Prediction},
author={Bowen Shi and others},
booktitle={Proc. ICLR},
year={2022}
}

@ARTICLE{10214650,
  author={Wang, Zhong-Qiu and others},
  journal={IEEE Trans. Audio, Speech, Language Process.}, 
  title={TF-GridNet: Integrating Full- and Sub-Band Modeling for Speech Separation}, 
  year={2023}}

@inproceedings{
xu2025tiger,
title={{TIGER}: Time-frequency Interleaved Gain Extraction and Reconstruction for Efficient Speech Separation},
author={Mohan Xu and others},
booktitle={Proc. ICLR},
year={2025},
url={https://openreview.net/forum?id=rzx3vcvlzj}
}

\end{document}